# The Common Difference Between MIMO With Other Antennas

MD Sirajul Huque  
*CSE & JNTU-Kakinada*

C Surekha  
*CSE & JNTU-Anantapur*

S Pavan Kumar Reddy  
*CSE & JNTU-Anantapur*

Vidhisha Yadav  
*CSE & JNTU-Anantapur*

*Abstract*— In past 802.11 systems there is a single Radio Frequency (RF) chain on the Wi-Fi device. Multiple antennas use the same hardware to process the radio signal. So only one antenna can transmit or receive at a time as all radio signals need to go through the single RF chain. In MIMO there can be a separate RF chain for each antenna allowing multiple RF chains to coexist. MIMO technology has attracted attention in wireless communications, because it offers significant increases in data throughput and link range without additional bandwidth or increased transmit power. It achieves this goal by spreading the same total transmit power over the antennas to achieve an array gain that improves the spectral efficiency (more bits per second per hertz of bandwidth) or to achieve a diversity gain that improves the link reliability. Multiple Input/Multiple Output (MIMO) is an area of intense development in the wireless industry because it delivers profound gains in range, throughput and reliability. As a result, manufacturers of wireless local area network (WLAN), wireless metropolitan area network (WMAN), and mobile phone equipment are embracing MIMO technology. In this paper we are interested to compare the MIMO Antenna functions with traditional Antenna functions. And we take an example of IRT for illustration.

*Keywords*— MIMO, WLAN, IRT, Spatial Multiplexing, Precoding, WiMAX.

## 1. INTRODUCTION MIMO

The multiple-input and multiple-output, or MIMO (commonly pronounced my-moh or me-moh), is uses the multiple antennas at both the sides like sender side and receiver side to improve communication performance. It is the one of smart antenna technology available in several forms. The terminologies input and out are belongs to the radio channel carrying the signal, not for the devices having antennas.

The MIMO technology is the one of interested technology in wireless communications, because it significantly increases the data throughput and link range too without additional bandwidth or increased transmit power. It goal is achieved due to by exploring the same by transmitting by power over the antennas to achieve an array gain that improves the spectral efficiency or to achieve a diversity gain that improves the link reliability. By these properties, MIMO is an important part of modern wireless communication technological standards such as IEEE 802.11n (Wi-Fi), 4G,3GPP Long Term Evolution, WiMAX and HSPA+.

The MIMO is sub-divided into three main categories, precoding, spatial multiplexing or SM, and diversity coding.

**Precoding** is multi-stream beam forming, in the narrowest definition. In general, it will be stated as spatial processing that occurs at the transmitter side. In (single-layer) beam forming, the same signal is emitted from each of the transmit antennas with appropriate phase (and sometimes gain) weighting such that the signal power is maximized at the receiver input. The benefits of beam forming are to increase the received signal gain, by making signals emitted from different antennas add up constructively, and to reduce the multipath fading effect. In the absence of scattering, beam forming results in a well defined directional pattern, but in typical cellular conventional beams are not a good analogy. When the receiver has multiple antennas, the transmit beam forming cannot simultaneously maximize the signal level at all of the receive antennas, and precoding with multiple streams is used. Note that precoding requires knowledge of channel state information (CSI) at the transmitter.

**1.1 Functions of MIMO**

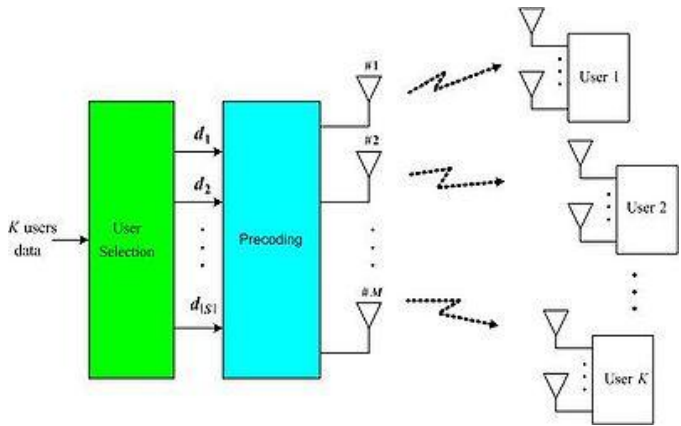

Fig 1 Precoding of MIMO technique

**In the spatial multiplexing** it is required that MIMO antenna configuration. In the spatial multiplexing, high rate signal will be split into multiple lower rate streams and each stream is transferred from a different antenna in the same frequency channel. After receiving these signals by the receiver antenna array with different spatial signatures, the receiver will separate these streams into parallel channels. Spatial multiplexing is the powerful technique to increase channel capacity at a higher signal-to-noise ratios (SNR). In (MIMO, wiki) the maximum number of spatial streams is limited by the less number of antennas at both ends of the transmitter or receiver. Spatial multiplexing may use with or without transmitting channel knowledge. Spatial multiplexing may also be used for simultaneous transmission of signals to multiple receivers, known asspace-division multiple accesses.

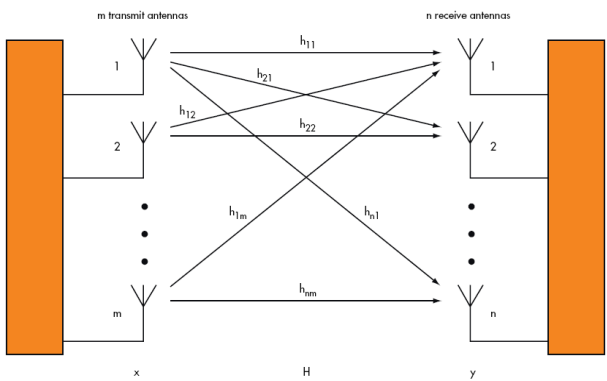

Fig 2 MIMO Spatial Multiplexing

**Diversity Coding** will be used when the transmitter doesn't have the channel knowledge. In this method, a single stream line is transmitted, but the signal is coded using space-time coding technique. The signal is emitted with full or orthogonal coding from each transmitted antennas. Diversity coding acts as the independent fading in the multiple antenna links to enhance signal diversity. Because there is no channel knowledge, there is no beam forming or array gain from diversity coding. For example, ZTE's WiMAX system provides higher spectrum utilization through favorable combination of MIMO and OFDMA techniques (MIMO, Sydaap). OFDMA has multiple orthogonal sub-carriers that increase spectrum efficiency, but the Cycle Prefix (CP) inserted at the end of each symbol may to some extent reduce transmission efficiency. When combined with the MIMO technique, the system can further increase spectral efficiency without expanding the bandwidth. MIMO can improve wireless communications in two ways: transmit diversity and spatial multiplexing. By using multiple paths between the transmitter and receiver antennas, the transmit diversity mechanism effectively reduces the Bit Error Rate (BER), improving robustness of the WiMAX system (see Figure 3). Different from the transmit diversity mechanism, the spatial multiplexing scheme transmits different information via different antennas and obtains spatial multiplexing gains, thus significantly increasing system capacity and spectrum utilization.

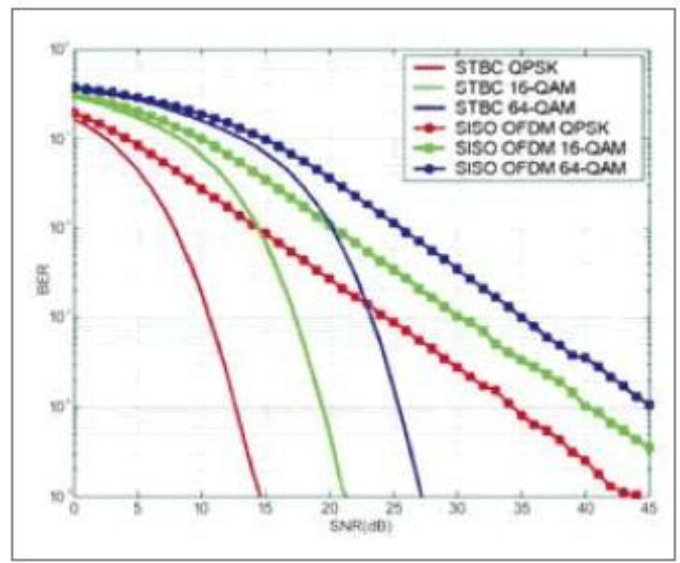

Fig 3 ZTE's MIMO uses three lines for Diversity coding

From the machine manufacturers view the product is used the "MIMO" concept to mean that it is "quicker, better, further", we often encounter that is inaccurate and even incorrect information and description as what actually an MIMO means. It is quite different, and this is unfortunately referred to channel bonding as MIMO. Channel bonding will increase the data rate by merging of several channels, but this occurs at the expenditure of the spectrum. Here "multiple channels" is often interchanged with "multiple antennas".

## 1.2 Receive Diversity and Maximum Ratio/Receive Combining

In *Switched Receive Diversity the* spatially separated receiving antennas will receive an emitted signal. A corresponding receiver switch will ensures that each antenna signal is utilized with the highest signal-to-noise ratio (SNR) for the actual data analysis.

In *Maximum Ratio Combining (MRC)* amplitudes and phases of the data signals received are adjusted with the help of digital signal processing in such a way that signal addition leads to gains in the S/N ratio and hence to a better bit error ratio (see Figure 1). Both are tested and used in multiplicity of transmission system but they should never be interchanged with MIMO.

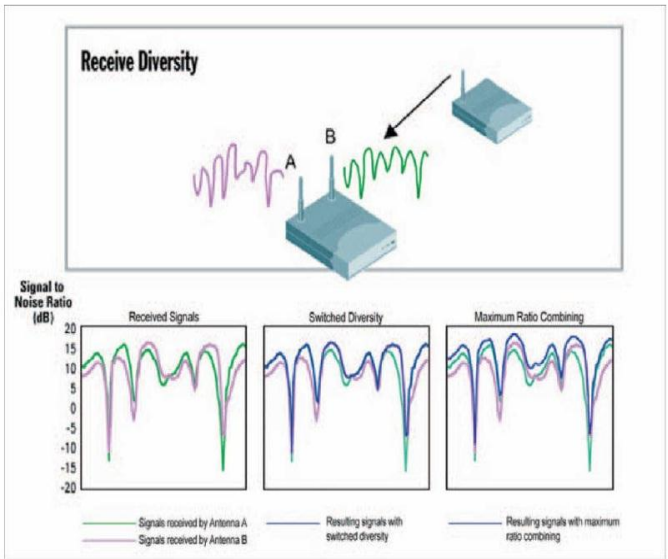

Figure 4: Comparison: Switched Receive Diversity and Maximum Ratio (Receive) Combining

## 1.3 Adaptive Antenna Systems – Beam forming

It is also a misleading that to refer a technique that is known as *Adaptive Antenna Systems (AAS)*, *Beam forming* and *Beam steering* just as MIMO technique. Several individual antenna elements are used to the sender and/or receiver side (Figure 2) but in terms of antennas they form **one** sending, one receiving antenna whose antenna lobe can be electronically adjusted. In contrast to **True MIMO**, the **same** data stream is sent through the transmission channel in all times.

True MIMO acts as main; therefore this technique is confined that these conceptual interpretations of MIMO. However, the combination of AAS/beam forming and True MIMO or even MRC is possible and also practical.

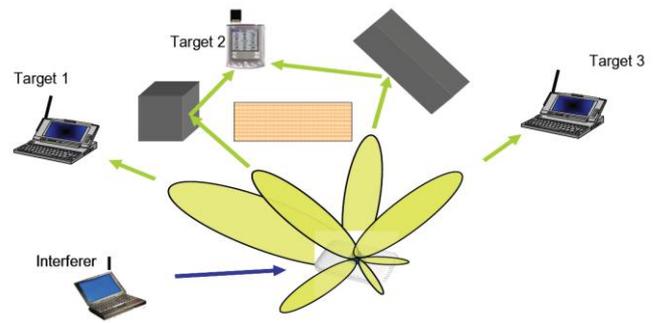

Figure 5: Adaptive Antenna Systems (AAS) can distribute the transmission energy in different ways in space with the help of several differently aligned antennas through so-called beam forming

## 1.4 Send-side Algorithm: Multilayer Systems

The concept of multilayer space-time signal processing was introduced in 1996 with the so-called Bell *Labs Layered Space-Time* procedure (BLAST procedure) [2, 5, and 6]. Here the data stream to be transmitted is divided into smaller packets. These individual data streams, called layers, are now emitted in encoded and modulated form through multiple antennas. In Figure 3a each layer consists of eight symbols, where two symbols of one layer are sent by one antenna. Due to its diagonal layout this procedure is called D-BLAST (diagonal BLAST).

Figure 3b shows the so-called V-BLAST (vertical BLAST) procedure, which tracks a division of the entire data stream in layers, the number of which coincides with the number of sending antennas. This spatially separated transmission type is called *Spatial Multiplexing* or *Space-Division Multiplexing* (*SM*).

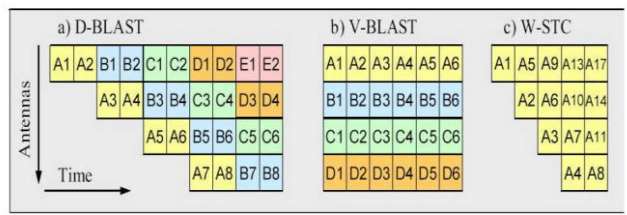

Figure 6: Layout of code words on the sender side in a) D-BLAST b) V-BLAST with four sending antennas

## 1.5 MIMO OFDM

OFDM is used in numerous wireless transmission standards nowadays (DAB, DVB-T, WiMAX IEEE 802.16, ADSL, WLAN IEEE 802.11a/g, Home Plug AV or DS2 200 aka "Home Bone"). The OFDM modulation transforms a broadband, frequency-selective channel into a multiplicity of parallel narrow-band single channels. A guard interval (called Cyclic Prefix CP) is inserted between the individual symbols. This guard interval must be temporally long enough to compensate for jitter in the transmission channel. Transmitted OFDM symbols experience different delays through the transmission channel. The variation of these delays at the receiving location is called jitter.

The appearance of *inter-symbol interference (ISI)* can thus be prevented. It has been shown in [8] that OFDM can be favourably combined with multiple antennas on the sending side as well as the receiving side to increase diversity gain and/or transmission capacity in time-varying and frequency-selective channels. The result is the MIMO OFDM systems now crowding the market.

In [1] a differentiation is made between Spatial Multiplexing MIMO (SM-MIMO) transmission systems with *per-stream coding (PSC)* and those with *per-antenna coding (PAC)*. In PSC the entire data stream is first encoded, the code bits are scrambled and then divided into parallel data streams according to the number of antennas. In PAC the encoding is done per sending antenna by all sub carriers.

### 1.6 Algorithms on the Receiving Side

The receive signal consists of the linear overlapping of the transmitted layers (Figure 4. therefore it is spoken of as interlayer interference (ILI). Without the corresponding detection methods, as compared with each other on the receiver side in [1], the transmitted information can no longer be directly evaluated. In particular, it is apparent that optimal detection methods are accompanied by rather high computing costs and the number of calculation steps increases significantly with the number of sending and receiving antennas. To reduce these huge computing costs, that is, to be able to implement SM-MIMO systems more efficiently, a sub-optimal detection method has been and is being sought for and worked on.

The complexity of the Maximum Likelihood (ML) detector grows exponentially with the size of the signal constellation, and this motivates the use of simpler suboptimum detectors in practical applications. Among those are [12]:

- ❖ Zero-forcing (ZF) detectors, which invert the channel matrix. The ZF receiver has a very small complexity that does not depend on the modulation. However, it does not exploit completely the system diversity and suffers from bad performance at low SNR.
- ❖ Minimum mean-square error (MMSE) detectors, which reduce the combined effect of interference between the parallel channels and additive noise. The MMSE receiver slightly improves the performance of the ZF receiver, but it requires knowledge of the SNR, which can be impractical. Besides, it does not exploit completely the channel diversity either.
- ❖ Decision-feedback receivers, which make a decision on one of the symbols and subtract its interference on the other symbol based on that decision. These receivers offer improved performance when compared to ZF and MMSE receivers, but they are prone to error propagation and still lack optimality, which may lead to large performance losses.
- ❖ Sphere detectors, which reduce the number of symbol values used in the ML detector.

Note that this type of detectors may preserve optimality while reducing implementation complexity.

In (WiMAX, 2007) the case of signal conditions are excellent; the data rate is doubled, tripled or quadrupled depending on the number of antennas used in both the sides of transmitter and receiver. In practice, spatial multiplexing is often implemented using more antennas in receiver side compare to transmitter antennas (WiFi 11n). In that case, the channel matrix is better conditioned and the performance degradation of suboptimal detectors (ZF, MMSE and Decision-Feedback) is reduced.

The first practical realizations of OFDM-based WLAN technique show that MIMO can now be made available to the mass market. With the new IEEE 802.11n standard, which should likely be accepted at the beginning of 2009, WiFi besides mobile WiMAX will become the technical trendsetters for other wireless mobile broadband systems [3, 7].

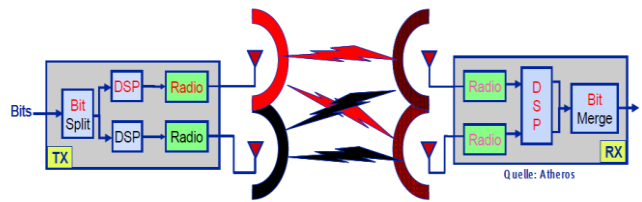

Figure 7: Superimposition of transmitted layers on the receiving antennas – spatial multiplexing

## 2. MIMO TECHNIQUES DEFINITIONS IN WIMAX SYSTEMS

WiMAX industry vendors and sources have created the host naming conventions to refer to multi antenna implementations which interprets the term MIMO more extensively – so it can be used to refer any multi-antenna techniques [11]. In (Ling Chen, 2007) The mobile WiMAX standard IEEE 802.16e-2005 includes two versions of MIMO techniques Matrix A and Matrix B. Space Time Block Coding (STBC) is the scheme referred as Matrix A and Spatial Multiplexing (SM-MIMO) is the scheme Matrix B.

MIMO Matrix A and MIMO Matrix B leverages multi-antenna operations at the base station and the end-user device. Matrix A and Matrix B with two antenna receivers is a required Wave 2 WiMAX Forum certification feature for

WiMAX devices and will be a supported capability in the broad pool of certified equipment.

Matrix A stands for coverage means:
- Will increase the radius of the cell
- provide better throughput for subscribers that are difficult to reach for terminals which already experience good signal conditions
- Matrix A has the benefit that higher order modulation will be used and fewer error correction bits are necessary which in turn increases transmission speeds to that subscriber.

Matrix B stands for capacity increase.

Collaborative Uplink MIMO is an additional MIMO technique considered by WiMAX Wireless LAN vendors to increase the spectral efficiency and capacity of the uplink communications path (Shamik Mukherjee, 2010). A practical realization of this technique would allows you, two separate end user WiMAX devices, each having a single transmit line-up, to utilize the same frequency allocation to communicate with the dual antenna WiMAX base station. This technique cans effectively double the uplink capacity of the WiMAX system.

It was proven that the best MIMO scheme to use in practice depends on the channel SNR and the required throughput as well as on other considerations such as the interference cancellation capability. Dedicated mechanisms such as the Fast Feedback Channel have been incorporated specifically in the WiMAX standard for the purpose of doing link adaptation. The best way in future to handle the MIMO schemes is to add the MIMO dimension to modulation and coding, and to select the best MIMO/Modulation/Coding combination through link adaptation [12].

Figure 8 depicts the 10 useful combinations for link adaptation over a pedestrian channel.

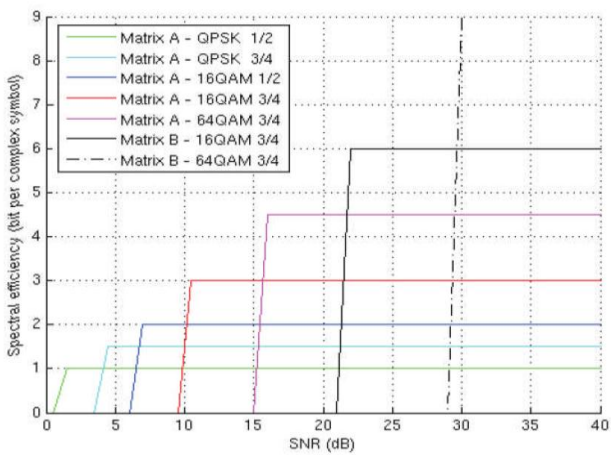

Figure 8: Operating SNR thresholds for adaptive modulation, coding and MIMO combinations source: Sequins Communications

## 3. MODULATION AND CODING IN IEEE 802.11N

The manufacturing union *Enhanced Wireless Consortium (EWC)* adopted its own proprietary specification in the middle of October 2005. The objective was to accelerate 802.11n standardization. The IEEE 802.11n working group adopted the innovations proposed by EWC into the draft standard, the "draft-n", and decided to prescribe them as *mandatory* for the definitive standard. Table 1 shows the numerous variants possible with adaptive modulation and coding. Beside MIMO, however, there are three further innovations in the physical layer (PHY) which can, with limited scope, contribute to the increase in data throughput or range.

- 20 MHz and 40 MHz are now possible as channel bandwidths. In a similar way as with the above-mentioned Channel Bonding, the data throughput here increases fully at the expense of the spectrum used.
- Through new and rapid chip techniques it has become possible to halve the cyclic prefix (guard interval GI), which is inserted between the individual OFDM symbols during transmission to prevent inter-symbol interference, from 800 ns to 400 ns. The last double column shows the corresponding gains in link rates (in comparison to the penultimate column).
- In the meantime, the proposal has been able to increase the number of OFDM sub carriers used in the data transmission from 48 to 52 (HT-OFDM High Throughput). This is why the link rate (gross data rate) increases from 54 Mbps to 58.5 Mbps. The real high-tech innovation, however, is to be found in the application of several "spatial streams" through the introduction of MIMO (Table 1, column 2: Number of spatial streams). For the first time, this makes it possible for spatially separate and different data to be sent on the same channel (the same centre frequency), which can be equated to the transfer from "Shared Ethernet" to "Switched Ethernet" through micro-segmentation.

For MIMO to achieve optimal improvement in the actual net throughput, upgrades had to be made in the MAC layer as well. In addition to shorter inter-frame spaces (IFS), it is now possible to combine frames and thus acknowledge them as blocks. To increase the throughput, the data quota of a single MAC frame – and thus the duration of user data transmission compared to the fixed times of the inter-frame spaces and the acknowledge packet – should be as large as possible.

| Bits 0-6 in HT-SIG1 (MCS index) | Number of spatial streams | Modulation | Coding rate | $N_{ES}$ | | $N_{SD}$ | | $N_{CBPS}$ | | GI = 800ns | | GI = 400ns | |
|---|---|---|---|---|---|---|---|---|---|---|---|---|---|
| | | | | 20 | 40 | 20 | 40 | 20MHz | 40MHz | Rate in 20MHz | Rate in 40MHz | Rate in 20MHz | Rate in 40MHz |
| 0 | 1 | BPSK | ½ | 1 | 1 | 52 | 108 | 52 | 108 | 6.5 | 13.5 | 7 2/9 | 15 |
| 1 | 1 | QPSK | ½ | 1 | 1 | 52 | 108 | 104 | 216 | 13 | 27 | 14 4/9 | 30 |
| 2 | 1 | QPSK | ¾ | 1 | 1 | 52 | 108 | 104 | 216 | 19.5 | 40.5 | 21 2/3 | 45 |
| 3 | 1 | 16-QAM | ½ | 1 | 1 | 52 | 108 | 208 | 432 | 26 | 54 | 28 8/9 | 60 |
| 4 | 1 | 16-QAM | ¾ | 1 | 1 | 52 | 108 | 208 | 432 | 39 | 81 | 43 1/3 | 90 |
| 5 | 1 | 64-QAM | ⅔ | 1 | 1 | 52 | 108 | 312 | 648 | 52 | 108 | 57 7/9 | 120 |
| 6 | 1 | 64-QAM | ¾ | 1 | 1 | 52 | 108 | 312 | 648 | 58.5 | 121.5 | 65 | 135 |
| 7 | 1 | 64-QAM | 5/6 | 1 | 1 | 52 | 108 | 312 | 648 | 65 | 135 | 72 2/9 | 150 |
| 8 | 2 | BPSK | ½ | 1 | 1 | 52 | 108 | 104 | 216 | 13 | 27 | 14 4/9 | 30 |
| 9 | 2 | QPSK | ½ | 1 | 1 | 52 | 108 | 208 | 432 | 26 | 54 | 28 8/9 | 60 |
| 10 | 2 | QPSK | ¾ | 1 | 1 | 52 | 108 | 208 | 432 | 39 | 81 | 43 1/3 | 90 |
| 11 | 2 | 16-QAM | ½ | 1 | 1 | 52 | 108 | 416 | 864 | 52 | 108 | 57 7/9 | 120 |
| 12 | 2 | 16-QAM | ¾ | 1 | 1 | 52 | 108 | 416 | 864 | 78 | 162 | 86 2/3 | 180 |
| 13 | 2 | 64-QAM | ⅔ | 1 | 1 | 52 | 108 | 624 | 1296 | 104 | 216 | 115 5/9 | 240 |
| 14 | 2 | 64-QAM | ¾ | 1 | 1 | 52 | 108 | 624 | 1296 | 117 | 243 | 130 | 270 |
| 15 | 2 | 64-QAM | 5/6 | 1 | 1 | 52 | 108 | 624 | 1296 | 130 | 270 | 144 4/9 | 300 |
| 16 | 3 | BPSK | ½ | 2 | 2 | 52 | 108 | 156 | 324 | 19.50 | 40.50 | 21.67 | 45.00 |
| 17 | 3 | QPSK | ½ | 2 | 2 | 52 | 108 | 312 | 648 | 39.00 | 81.00 | 43.33 | 90.00 |
| 18 | 3 | QPSK | ¾ | 2 | 2 | 52 | 108 | 312 | 648 | 58.50 | 121.50 | 65.00 | 135.00 |
| 19 | 3 | 16-QAM | ½ | 2 | 2 | 52 | 108 | 624 | 1296 | 78.00 | 162.00 | 86.67 | 180.00 |
| 20 | 3 | 16-QAM | ¾ | 2 | 2 | 52 | 108 | 624 | 1296 | 117.00 | 243.00 | 130.00 | 270.00 |
| 21 | 3 | 64-QAM | ⅔ | 2 | 2 | 52 | 108 | 936 | 1944 | 156.00 | 324.00 | 173.33 | 360.00 |
| 22 | 3 | 64-QAM | ¾ | 2 | 2 | 52 | 108 | 936 | 1944 | 175.50 | 364.50 | 195.00 | 405.00 |
| 23 | 3 | 64-QAM | 5/6 | 2 | 2 | 52 | 108 | 936 | 1944 | 195.00 | 405.00 | 216.67 | 450.00 |
| 24 | 4 | BPSK | ½ | 2 | 2 | 52 | 108 | 208 | 432 | 26.00 | 54.00 | 28.89 | 60.00 |
| 25 | 4 | QPSK | ½ | 2 | 2 | 52 | 108 | 416 | 864 | 52.00 | 108.00 | 57.78 | 120.00 |
| 26 | 4 | QPSK | ¾ | 2 | 2 | 52 | 108 | 416 | 864 | 78.00 | 162.00 | 86.67 | 180.00 |
| 27 | 4 | 16-QAM | ½ | 2 | 2 | 52 | 108 | 832 | 1728 | 104.00 | 216.00 | 115.56 | 240.00 |
| 28 | 4 | 16-QAM | ¾ | 2 | 2 | 52 | 108 | 832 | 1728 | 156.00 | 324.00 | 173.33 | 360.00 |
| 29 | 4 | 64-QAM | ⅔ | 2 | 2 | 52 | 108 | 1248 | 2592 | 208.00 | 432.00 | 231.11 | 480.00 |
| 30 | 4 | 64-QAM | ¾ | 2 | 2 | 52 | 108 | 1248 | 2592 | 234.00 | 486.00 | 260.00 | 540.00 |
| 31 | 4 | 64-QAM | 5/6 | 2 | 2 | 52 | 108 | 1248 | 2592 | 260.00 | 540.00 | 288.89 | 600.00 |
| 32 | 1 | BPSK | ½ | 1 | 1 | | | 48 | | 48 | | 6 | 6.67 |

Table 1: Modulation and coding scheme (MCS) in the new standard IEEE 802.11n. NCBPS: Number of coded bits per symbol - NES: Number of FEC encoders – NSD: Number of data sub carriers - GI: Size of guard interval.

## 4. COMPARATIVE MEASUREMENTS

For example The Institut für Rundfunktechnik (IRT) has conducted and is conducting performance measurements on the latest pre-n and draft-n products available on the market. A real comparison of the new techniques with the existing 802.11g standard can be made only if all measurements are performed at channel bandwidths of 20 MHz – that is without channel bonding. Unfortunately, several implementations do not allow the restriction of the sending spectrum to one WLAN channel. In selecting the sending location and the different receiving locations, it was ensured that all site situations that occur in practice were mapped against each other as far as possible. The measurements were performed in a two-dimensional area, in other words on one floor of the building. For the duration of the measurement, all other access points in the measurement range were switched off or converted to sending channels whose spectral masks do not overlap in any way with the spectral masks of the measurement channel.

The location of the MIMO router is shown in green. There are in total seven locations and a measurement laptop with the corresponding MIMO PC card was placed at each one. The separating walls of the individual rooms, as can be seen from the floor plan, consist of brickwork about 15 centimeters thick and four fire protecting walls constructed out of poured concrete 30 centimeters thick. The damping of these fire barriers thus corresponds to the damping provided by conventional concrete ceilings or floors or two layers of thermal double glazing [10].

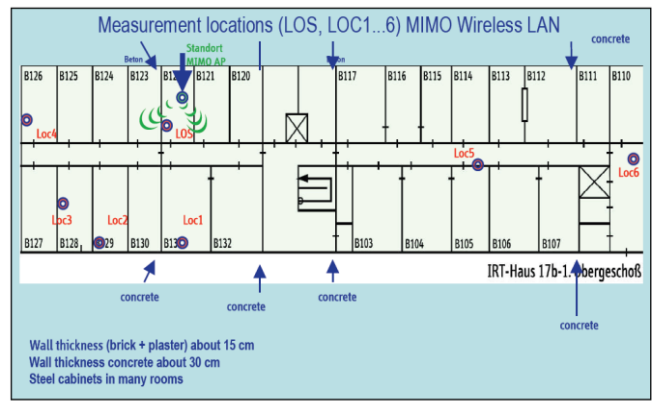

Figure 9: The WLAN measurement environment at the IRT on the upper floor of building 17B.

### 4.1 Measurement Layout

The MIMO-AP was placed at a height of about 1.8 meters at the spatial position that can be seen in Figure 9. The measurement locations of the MIMO laptops were each at about one meter above the floor. A measurement location was located in the sending room at a three-meter distance with direct line of sight (LOS); all other six measurement locations were non-line-of-sight locations (NLOS).

A view of the building plan shows that wave propagation in the 2.4 GHz range within the floor is marked with manifold dispersions and reflections. Since the send signal reaches the recipient at NLOS in the most varied and indirect ways, this automatically leads to phase displacements and variable weakening of the individual signals. The overlapping of the individual signals at the receiving location (Figure 10) leads to signal fall-offs that fluctuate over time and frequency (multipath fading). However, these negative effects of multipath fading in SISO systems can be turned to a positive by using spatial multiplexing in MIMO OFDM systems.

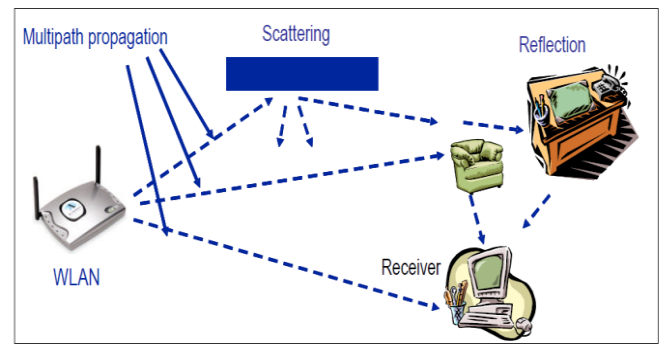

Figure 10: Principle of multipath propagation

Figure 11 shows the measurement equipment used. The measurement software "IxChariot from IXIA was used. This software is designed for professional performance and QoS measurement based on protocols such as TCP or UDP.

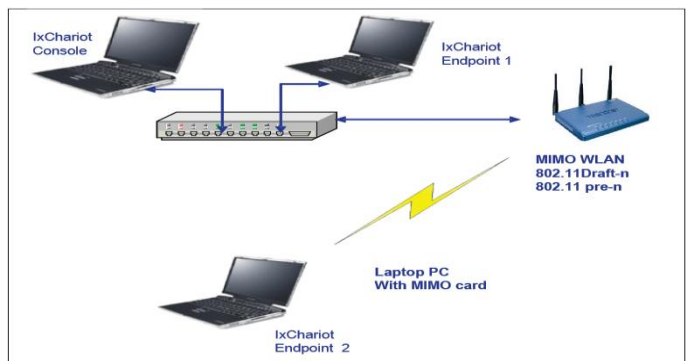

Figure 11: The measurement equipment for the MIMO measurements at IRT consists of a Gigabit Ethernet switch, to which the WLAN access point to be tested is connected, a laptop with the counterpart to be tested – the WLAN client card, as well as another laptop on which the measurement software is running.

Channel 13 was selected for the measurement. However, for purposes of comparison each product was randomly tested again on channel 6. There were no noteworthy changes in the results. The MIMO PC card to be measured was inserted into a Siemens/Fujitsu Lifebook (model S7010D S-Series Centrino). Random measurements were repeated and compared here as well, using a Samsung X10 laptop. The results of the measurement were not dependent on the laptop used here either. Windows XP Professional with current patches was used as the operating system in all PCs.

The driver and utility software for each individual MIMO PC card was installed on the respective measurement laptop from the accompanying software CD and deinstalled after completion of the measurements. Prior to the draft-n measurements conducted in November/December 2006, the then most current firmware for the MIMO-WLAN router and the most current driver and utilities for the MIMO client cards were installed from the respective manufacturer websites.

TCP performance measurements were conducted at all seven locations. The transmission direction was "downstream", in accordance with the expected applications, that is, directed from the access point to the client. Each individual measurement lasted three minutes and each one was performed three times. Bar graphs show the average values from the three measurements in each location (and this is an average of 9 minutes length of time which means IRT examined very carefully). By using the newest version of the measurement software, IxChariot (V6.4), test transmissions could be done on the pre-n, draft-n 1.0/2.0 products themselves especially using the "High Performance Throughput" script which is provided by IxChariot. This script is perfectly suited for doing comparative measurements of maximum TCP performance over certain kinds of networks. With predefined scripts from IxChariot also IPTV streaming, MPEG2 video streaming and VoIP measurements can be done and highlighted according to quality criteria such as delay and jitter. Several SD- and HDTV streaming measurements were carried out as well at the IRT using data rates reaching from 4Mbps up to 20Mbps per stream.

## 5. CONCLUSIONS

In present and future the wireless communications systems MIMO-OFDM has become an eminent technology of wireless and mobile communications. Its potential to increase spectral efficiency has not been reached by any other technique before. In addition to increasing spectral efficiency, MIMO can also be used to reduce transmitting power while keeping coverage areas constant. The use of MIMO technique in future transmission systems for broadcasting, multicasting and unicasting represents real business logic also for broadcasting corporations because of the possible reduction in transmission stations. The measurements conducted at IRT also show, as described in [9], very good MIMO properties with line of sight in the indoor area. What is important is that adequate multipath scattering and thus "gain-bringing multiple reception" emerge. Something that has been regarded as an annoyance in transmission in radio technique for a hundred years has become an advantage to users through the smart application of physics and mathematics. Because signal transmission is always analogue – digital bits come into being only through calculation.

## References

[1] Dirk Wübben, „Effiziente Detektionsverfahren für Multilayer-MIMO-Systeme", Dissertation Universität Bremen 2005
[2] Working Group 8, „Space-Time signal Processing and MIMO Systems", White Paper 2004
[3] Datacomm Research Company, "Using MIMO-OFDM Technology to boost Wireless LAN Performance today", White Paper 2005
[4] IEEE, "Broadband MIMO-OFDM Wireless Communications", Invited Paper 2004
[5] Helmut Bölcskei, "Principles of MIMO-OFDM Wireless Systems", Swiss Federal Institute of Technology (ETH) Zürich
[6] Helmut Bölcskei, "MIMO-OFDM Wireless Systems: Basics, Perspectives, and Challenges", Communication Technology Laboratory ETH Zürich 2007
[7] Heiko Schmidt, "OFDM für die drahtlose Datenübertragung innerhalb von Gebäuden", Dissertation Universität Bremen 2001
[8] Xiang, Waters, Bratt, Barry, Walkenhorst, „Implementation and Experimental Results of a Three-Transmitter Three-Receiver OFDM/BLAST Testbed", IEEE Communicatin Magazine 2004
[9] Wojciech Kuropatwinski-Kaiser, "MIMO-Demonstrator basierend auf GSM-Komponenten"
[10] Bölcskei, Gespert, Papadias, van der Veen, "Space-Time Wireless Systems"
[11] White Paper Motorola, "A practical guide to WiMAX Antennas: MIMO and Beamforming Technical Overview", 2006
[12] White Paper Sequans Communications, "MIMO Techniques for Mobile WiMAX Systems", 2006
[13] www.tomsnetworking.de
[14] http://www.irt.de/de/themengebiete/digitale-netze/mimo-ofdm.html (blue technical report - german only - No. B 197/2006)
[15] Chetan Sharma Consulting, "US Wireless Data Market Q4 2009 and 2009 Update," available on the web at: http://www.chetansharma.com/US%20Wireless%20Market%20Q4%202009%20Update%20-%20Mar%202010%20-
 %20Chetan%20Sharma%20Consulting.ppt#261,1 on March 3, 2010.

[16] Information Content estimate from www.wolframalpha.com, March 3, 2010.
[17] "MIMO Transmission Schemes for LTE and HSPA Networks," 3G Americas Whitepaper, June 2009.
[18] "Vocabulary of 3GPP Specifications," TR21-905, 3GPP, version 8.8.0, March 2009.
[19] Rysavy Research, "HSPAto LTE-Advanced: 3GPP Broadband Evolution to IMT-Advanced (4G)," 3G Americas White Paper, September 2009.
[20] Jonas Karlsson and Mathias Riback, "Initial Field Performance Measurements of LTE," Ericsson Review No. 3, 2008.
[21] Krishna Balachandran, Qi Bi, Ashok Rudrapatna, James Seymour, Soni, and Andreas Weber, "Performance Assessment of Next-Generation Wireless Mobile Systems," Bell Labs Technical Journal 2009 vol. 13:4, pages 35-58, http://www3.interscience.wiley.com/journal/122211215/abstract last accessed on March 30, 2010.
[22] A.M.D. Turkmani, A.A. Arowojolu, P.A. Jefford, and C.J. Kellett, "An Experimental Evaluation of the Performance of Two-Branch Space and Polarization Diversity Schemes at 1800 MHz," IEEE 0018-9545/94404,00, 1995.
[23] Brian S. Collins, "The Effect of Imperfect Antenna Cross-Polar Performance on the Diversity Gain of a Polarization-Diversity Receiving System," Microwave Journal, April 2000.
[24] R. Bhagavatula, R. W. Heath, Jr., and K. Linehan, "Performance Evaluation of MIMO Base Station Antenna Designs," Antenna Systems and Technology Magazine, vol. 11, no. 6, pp. 14 -17, Nov/Dec. 2008.
[25] Lawrence M. Drabeck, Michael J. Flanagan, Jayanthi Srinivasan, William M. MacDonald, Georg Hampel, and Alvaro Diaz, "Network Optimization Trials of a Vendor-Independent Methodology Using the Ocelot™ Tool," Bell Labs Technical Journal 9(4), 49–66 (2005) © 2005
[26] "Final Report on Semi-Smart Antenna Technology Project," Ofcom Contract No. 830000081, Document No. 830000081/04, July 2006.
[27] M. Petersen, et al., "Automatic Antenna Tilt Control for Capacity Enhancements in UMTS FDD," IEEE 2004, 0-7803-8521-7/04, 2004.
[28] Mano D. Judd, Thomas D. Monte, Donald G. Jackson, and Greg S. Maca, "Tranceive Distributed Antenna Systems," U.S. Patent No. US 6,597,325 B2, July 22, 2003.
[29] "Active Antenna Arrays, Small-Footprint, Scalable RF Solutions for Base Stations," Bell Labs Research Project, Alcatel-Lucent Innovation Days, December 2008. Available on line at http://innovationdays.alcatel-lucent.com/2008/documents/Active%20Antenna%20Array.pdf.

## Author's Profile

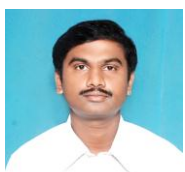

Mr.MD SIRAJUL HUQUE received his B.Tech in Computer Science & Engg from Nimra College of Engg & tech,JNTU Hyderabad.Pursuing M.tech in Computer science & engg from NCET,Vijayawada affliated to JNTUK. he has 5+ years of teaching Experience.He is working as Assistant Professor in St.Anns College of Engg & Tech,Chirala Affliated to JNTUK.

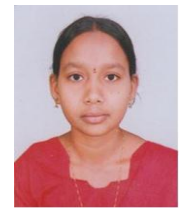

Ms.C.Surekha received her B.Tech in CSE from Vaagdevi Institute of Technology and Sciences and doing her M.Tech in Computer Science & Engineering from VITS in JNTU – Anantapur. She has 6 months of teaching experience.

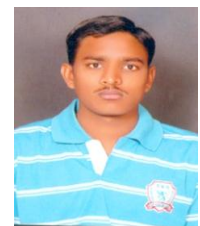

**Mr. S. Pavan Kumar Reddy** received his B.Tech (CSE) from Vaagdevi Institute of Technology and Sciences, JNTU-Anantapur, and pursuing M.Tech in Computer Science and Engineering from Vaagdevi Institute of Technology and Sciences, JNTU-Anantapur.

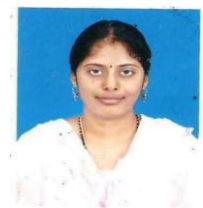

**Mrs Vidhisha Yadav** received her B.Tech (Biotechnology) and pursuing M.Tech in Computer Science and Engineering from Vaagdevi Institute of Technology and Sciences, JNTU-Anantapur.